\documentclass[aps,prl,twocolumn,groupedaddress,bibnotes]{revtex4}
\usepackage{graphicx,epsfig,subfigure,afterpage,bm}
\usepackage{afterpage,bm}
\usepackage{afterpage,bm,amsmath}

\newcommand{\be}{\begin{equation}}
\newcommand{\ee}{\end{equation}}
\newcommand{\bea}{\begin{eqnarray}}
\newcommand{\eea}{\end{eqnarray}}
\newcommand{\edot}{\dot{\epsilon}}
\newcommand{\edotbar}{\overline{\dot{\epsilon}}}
\newcommand{\epsbar}{\overline{\epsilon}}
\newcommand{\gdot}{\dot{\gamma}}

\newcommand{\vs}{{\it vs\/\;}}

\newcommand{\etal}{{\it et al.\/}}
\newcommand{\bw}{\begin{widetext}}
\newcommand{\ew}{\end{widetext}}

\newcommand{\Considere}{Consid\`{e}re\,\,}
\newcommand{\sig}{\tens{W}}
\newcommand{\Sig}{\tens{\Sigma}}
\newcommand{\tensile}{\sigma_{\rm E}}

\newcommand{\vecv}[1]{\bm{{#1}}}
\newcommand{\tens}[1]{\bm{{#1}}}
\newcommand{\nablu}{{\bf \nabla}}

\begin{document}

\title{Criterion for extensional necking instability in polymeric fluids}
\author{Suzanne M. Fielding}
\email{suzanne.fielding@durham.ac.uk}
\affiliation{Department of Physics, Durham University, Science Laboratories, South Road, Durham. DH1 3LE, U.K.}
\date{\today}
\begin{abstract} 
We study the linear instability with respect to necking of a filament
of polymeric fluid undergoing uniaxial extension.  Contrary to the
widely discussed \Considere criterion, we find the onset of
instability to relate closely to the onset of downward curvature in
the time (and so strain) evolution of the $zz$ component of the
molecular strain, for extension along the $z$ axis. In establishing
this result numerically across five of the most widely used models of
polymer rheology, and by analytical calculation, we argue it to apply
generically. Particularly emphasized is the importance of polymer
chain stretching in partially mitigating necking. We comment finally
on the relationship between necking and the shape of the underlying
steady state constitutive curve for homogeneous extension.
\end{abstract}
\pacs{ {a}, {b}.  }  \maketitle

Understanding the rheology (flow properties) of polymeric fluids is
central to their processing and performance. Many commercially
important flows are dominated by extensional components: fibre
spinning, film blowing, extrusion, and ink jet printing provide good
examples. In fluid dynamical terms, extensional flows cause material
elements to separate exponentially quickly and so subject the
underlying macromolecules to extreme stretching and
reorientation. They are thus highly sensitive to underlying molecular
details: linear \vs branched polymer chains, for example.  Indeed many
nonlinear flow features arise only in extension, which thus provides a
crucial benchmark for theories of polymer rheology.

In many polymeric fluids, a state of uniform flow becomes unstable
when the flow rate exceeds the rate $1/\tau$ on which the underlying
molecular structure relaxes~\cite{review}. In flows dominated by
shear, for example, the phenomenon of shear
banding~\cite{manneville2008} arises widely for shear rates $\gdot
\tau>O(1)$. In a linear stability analysis, the criterion for onset of
a shear banding instability is (usually) that the shear stress is a
decreasing function of shear rate.

\begin{figure}[b]
\vspace{-0.4cm}
  \includegraphics[width=4.4cm]{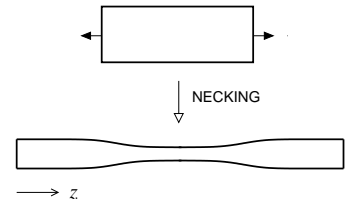}
\vspace{-0.2cm}
  \caption{Cartoon of necking.}
\label{fig:neck}
\end{figure}

In extensional rheology~\cite{reviews,reviews1} a common protocol is that of
filament stretching~\cite{TirtSrid93}, in which a cylindrical sample
of fluid is drawn out in length. Though the aim is to achieve uniform
extensional flow for benchmarking against theory, complications often
arise.  Particularly serious is the widespread observation of necking
instabilities (Fig.~\ref{fig:neck}), in both
experiment~\cite{cogswell1974,vinogradov1975c,inkson1999,mckinleyExp,lee2002,barroso,wang,rothstein}
and simulation~\cite{sizaire1997,hassager,yao1998a,bhat2008}. These
lead to heterogeneous deformation and even failure at only modest
strains: any small indentations in cross-sectional area become ever
more pronounced until the sample breaks. This hinders attempts to
characterise these fluids scientifically, and process them
commercially.

Although a long standing
problem~\cite{doi1979dcp,vincent1960,ide1977,pearson1982,lagnado1985,leonov1990,mckinley1999,olagunju,joshi,cromer2009,petrie2009,renardy2009b},
necking remains poorly understood. Crucially lacking is any reliable
criterion for its onset, of the same stature as that given above for
shear banding. Popularly discussed is the Consid\`{e}re
criterion~\cite{Considere}, which predicts necking if the tensile
force decreases with extensional strain: $dF/d\epsilon<0$.  But taking
$F$ to depend only on $\epsilon$ in this way assumes the flow purely
elastic: it cannot account for the dependence of necking on the strain
rate $\edot$ in these viscoelastic fluids; nor predict the rate at
which necking sets in.

Here we show that the \Considere criterion in fact does not apply in
most regimes of polymeric flow. By a linear stability analysis, we
demonstrate the onset of necking instead to relate closely to that of
downward curvature in the time (or equivalently strain) evolution of
the $zz$ component of the molecular conformation tensor, for
stretching at constant rate $\edot$ along the $z$ axis. In most
regimes, this further corresponds to downward curvature in the
evolution of the tensile stress. In establishing this result
numerically across five of the most widely used models of polymer
rheology, and by analytical calculation, we argue it to apply
generically. The other central contribution of this Letter is to
demonstrate the crucial role played by polymeric chain stretching in
partly mitigating necking.

We consider a polymeric fluid in inertialess flow, obeying the force
balance condition $0 = \nablu.\tens{T} = \nablu .(\Sig + 2\eta\vecv{D}
-P\tens{I})$.  The total stress $\tens{T}$ has a viscoelastic
contribution $\Sig$ from the polymer; a Newtonian contribution of
viscosity $\eta$ from the solvent; and an isotropic pressure field $P$
set by mass balance $\nablu.\vecv{v}=0$ for incompressible flow. Here
$\vecv{v}$ is the velocity field and $\tens{D}$ the symmetric part of
the velocity gradient tensor $\tens{\kappa}_{\alpha\beta}\equiv
\partial_\alpha v_\beta$.  We assume the polymeric stress
$\Sig=G_0[k(\sig)\sig-\tens{I}]$ with $G_0$ a modulus and $\sig$ a
dimensionless tensor characterising the conformation of the chainlike
polymer molecules. (In the simplest cartoon these are taken as
dumbbells with span $\vec{R}$ and $\tens{W}\propto \langle
\vec{R}\vec{R}\rangle$.) This has generalised dynamics
\be
\label{eqn:dynamics}
(\partial_t+\vecv{v}\cdot\nablu )\,\sig = \sig\cdot\tens{\kappa} + \tens{\kappa}^T\cdot\sig -\frac{1}{\tau}\tens{R}(\sig),
\ee
representing a competition between driving out of equilibrium by flow
(terms in $\tens{\kappa}$) and relaxation back to equilibrium
$\sig=\tens{I}$ on a timescale $\tau$.  

We perform calculations for five concrete functional choices for
$k(\sig)$ and $\tens{R}(\sig)$, each corresponding to a particular
model used widely in the literature~\cite{larson88}. The Oldroyd B
model has $k=1$, $\tens{R}=\sig-\tens{I}$. The Giesekus model differs
from this in having $\tens{R}=(\sig-\tens{I})+\alpha(\sig-\tens{I})^2$
with $0\le\alpha\le 1$; the Fene-P (in which we assume small $\delta$)
in having $\tens{R}=k(\sig)\sig-\tens{I}$ with $k=1/(1-\delta\,T)$ and
trace $T=W_{xx}+W_{yy}+W_{zz}$; the Rolie-Poly~\cite{likhtmangraham03}
$\tens{R}=\sig-\tens{I}+2(1-\sqrt{3/T})\left[\sig+\beta(T/3)^{\delta_{\rm
      RP}}(\sig-\tens{I})\right]\tau/\left[\tau_{\rm
    R}(1-fT/3)\right]$. Finite $\tau_{\rm R}$ in the Rolie-Poly model
allows stretching of the underlying polymer chains by the flow
field. Indeed for $\edot\tau_{\rm R}>1$ infinite stretch can develop
if $f=0$; $f>0$ restores finite stretch; $\tau_{\rm R}\to 0$ disallows
it entirely.

We use units in which $\tau=1$, $G_0=1$.  This leaves as parameters
the Newtonian viscosity $\eta$, which we take to zero; and $\alpha$
(Giesekus), $\delta$ (Fene-P), $\beta,\tau_{\rm R}, \delta_{\rm RP},f$
(Rolie-Poly) in which we use $\beta=0.0, \delta_{\rm RP}=-1/2$
following~\cite{likhtmangraham03}.

\begin{figure}[tb]
  \includegraphics[width=7.0cm]{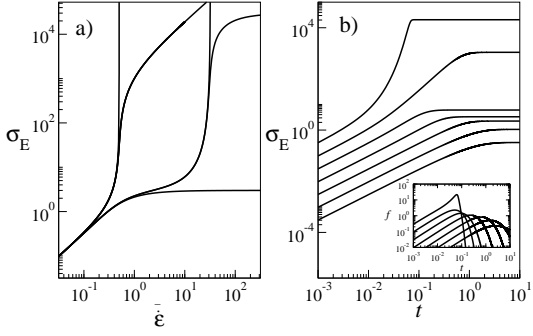}
\vspace{-0.3cm}
  \caption{a) Homogeneous constitutive curves for steady state
    uniaxial extension. Clockwise: Oldroyd B; Giesekus
    ($\alpha=0.001$); fene-P ($\delta=0.001$); stretch Rolie-Poly
    ($\tau_{\rm R}=0.0316$) with zero extensibility; the same with
    finite extensibility ($f=10^{-4}$); non-stretch
    Rolie-Poly. (Curves for Giesekus and fene-P indistinguishable.)
    b) Tensile stress $\tensile$ \vs time in startup of uniaxial
    extension in the stretch Rolie-Poly model with $\tau_{\rm
    R}=0.0316, f=10^{-4}$ for $\edot=10^n$, $n=-1,-1/2,0,1/2,1,3/2,2$
    (curves upwards). Inset: corresponding tensile force.}
\label{fig:base}
\end{figure}

We consider a long slender~\cite{denn1975} tube of fluid subject to
uniaxial extension along the $z$ axis. At any time $t$ it has cross
sectional area profile $A(z,t)$ and area-averaged fluid velocity
$V(z,t)$ in the $z$ direction, with extension rate
$\edot(z,t)=\partial_zV(z,t)$. The mass balance condition is then
$\partial_t A(z,t)=-\partial_z(AV)$, and force balance
$F(t)=A\left[k(T)(Z-X)+3\eta\edot\right] \equiv A\tensile$ (neglecting
surface tension).  This defines the tensile stress $\tensile(z,t)$
given force $F(t)$, which is uniform along the filament. $Z=W_{zz}$
and $X=\tfrac{1}{2}W_{xx}+\tfrac{1}{2}W_{yy}$ (so $Z+2X=T$) obey
\bea
\partial_tZ(z,t)+V\partial_zZ &=& 2\edot Z - R(Z,T)/\tau,\nonumber\\
\partial_tX(z,t)+V\partial_zX &=& - \edot X - R(X,T)/\tau,\label{eqn:components}
\eea
with $R(Z,T)=Z-1$ (Oldroyd B); $R(Z,T)=Z-1+\alpha(Z-1)^2$
(Giesekus); $R(Z,T)=k(T)Z-1$ (Fene-P);
$R(Z,T)=Z-1+2(1-\sqrt{\tfrac{3}{T}})\left[Z+\beta(\tfrac{T}{3})^{\delta_{\rm
RP}}(Z-1)\right]/\tau_{\rm R}(1-\tfrac{fT}{3})$ (Rolie-Poly).

Given a time-dependent separation of the sample ends that imposes a
global strain $\epsbar(t)$ averaged along the filament, we transform
to the affinely co-extending, co-thinning frame by defining
$u=ze^{-\epsbar(t)}$, $v(u,t)=V(z,t)e^{-\epsbar(t)}$ and
$a(u,t)=A(z,t)e^{\epsbar(t)}$.  We then have mass balance, force
balance, and  viscoelastic dynamics:
\vspace{-0.1cm}
\bea
\label{eqn:massEdotbar}
\partial_t a&=&-\partial_u\left[(v-\edotbar u)a\right],\\
0&=&\partial_u[a(k(T)(Z-X)+3\eta\edot)]=\partial_u[a\tensile],\\
\label{eqn:viscoEdotbar}
\partial_tZ&=&-(v-\edotbar u)\partial_uZ + 2\edot Z-R(Z,T)/\tau,\nonumber\\
\partial_tX&=&-(v-\edotbar u)\partial_uX-\edot X -R(X,T)/\tau.
\eea
\vspace{-0.2cm}

\vspace{-0.3cm}
For a constant global extension rate $\edotbar$ commenced at time
$t=0$, a homogeneous ``base state'' in which uniaxial extension is
(artificially) maintained uniformly along the filament, regardless of
whether it is in practice unstable to heterogeneous necking, is
prescribed by $\edot(u,t)=\edotbar$, $v(u,t)=u\edotbar$,
$a(u,t)=a_0=1$, and the homogeneous solutions $Z(t),X(t)$ of
Eqns.~\ref{eqn:viscoEdotbar} given $Z(0)=X(0)=1$.

In any regime of finite extensional viscosity this base state attains
as $t\to\infty$ a steady state in the co-thinning, co-extending
frame. (In the laboratory frame the sample exponentially extends and
thins.)  This is described by a homogeneous constitutive curve
$\tensile(\edot)$ of tensile stress \vs strain rate,
Fig.~\ref{fig:base}a. (For this uniform base state we use symbols
$\edotbar$ and $\edot$ interchangeably.)  The Oldroyd B model has
divergent viscosity for $\edot\to 1/2$. This divergence is avoided
(narrowly, for small $\alpha,\delta$) in Giesekus and Fene-P (which
reduce to Oldroyd B for $\alpha=0,\delta=0$). The stretch Rolie-Poly
model with $f=0$ likewise has divergent viscosity for $\edot \tau_{\rm
  R}\to 1$, avoided narrowly for small $f>0$ and entirely in the
non-stretch model $\tau_{\rm R}=0$.

The time-dependence of the tensile stress $\tensile$ in this
homogeneous base state as it evolves towards the steady state just
described is shown in Fig.~\ref{fig:base}b for the stretch Rolie-Poly
model with finite extensibility. The corresponding force
$F(t)=\tensile(t)A(t)$ (inset) displays a maximum due to the rise in
$\tensile$ after the inception of flow being later overcome by
exponentially declining area.

A maximum in force \vs time $t$ directly also signifies a maximum \vs
strain $\epsilon=\edot t$, given constant $\edot$. For a nonlinear
elastic solid, the \Considere criterion decrees such a maximum to
herald departure from uniform extension and onset of necking. We now
examine whether this criterion also applies to polymeric fluids, as
often suggested.

\begin{figure}[tb]
  \includegraphics[width=7.75cm]{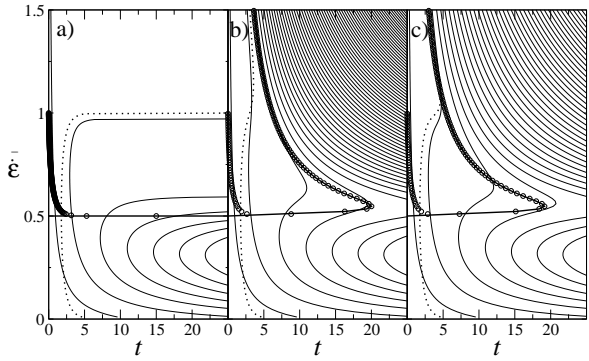}
\vspace{-0.35cm}
  \caption{Contour map for growth of fluctuations in area: $\delta
    a(t)/\delta a(0)=2^n$ for $n=1,2,3\cdots$ shown by thin solid
    lines from left to right. Dotted lines, open circles, thick solid
    lines show location respectively of $\dot{f}=0$,
    $\ddot{\tensile}=0$, $\ddot{Z}=0$. a) Oldroyd B, b) Giesekus
    ($\alpha=0.001$), c) Fene-P ($\delta=0.001$).}
\label{fig:OGF}
\end{figure}

To do so, we analyse the linear stability of the time-dependent
homogeneous uniaxially extending base state just described by
linearising in small heterogeneous fluctuations $\sum_q[\delta
  a_q(t),\delta \edot_q(t), \delta Z_q(t), \delta X_q(t)]\exp(iqu)$
about it. To first order in the amplitude of these we have
\be
\partial_t\left(\begin{matrix}
               \delta a\\
               0\\
               \delta Z\\
               \delta X\end{matrix}
 \right)_q=\left(\begin{matrix}
              0 & -1 & 0 & 0 \\
              \tensile & 3\eta & \partial_Z \tensile & \partial_X \tensile\\
              0 & 2Z & 2\edotbar-\frac{r}{\tau}-\frac{s}{\tau} & -\frac{2s}{\tau} \\
              0 & -X & -\frac{s}{\tau} & -\edotbar-\frac{r}{\tau}-\frac{2s}{\tau} \\
               \end{matrix}\right)
\left(\begin{matrix}
               \delta a\\
               \delta \edot\\
               \delta Z\\
               \delta X\end{matrix}
 \right)_q,\nonumber\label{eqn:linear}
\ee
in which $r=\partial_ZR(Z,T)$, $s=\partial_TR(Z,T)$. The
$q-$independence of this matrix suggests all spatial modes will grow
(or decay) equally, though we return below to discuss what shape can be
expected in practice.

\begin{figure}[tb]
  \includegraphics[width=7.5cm]{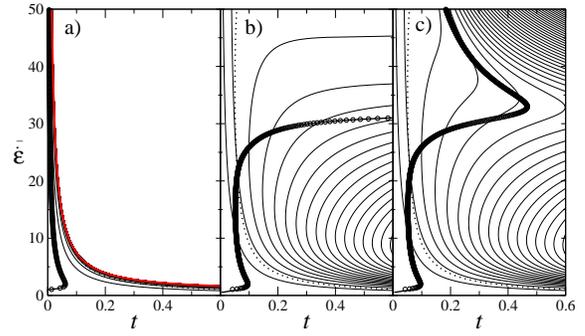}
\vspace{-0.35cm}
  \caption{Contour maps $\delta a(t)/\delta a(0)=2^n$ for
    $n=1,2,3\cdots$, thin solid lines left to right. Dotted lines,
    open circles, thick solid lines show location of $\dot{f}=0$,
    $\ddot{\tensile}=0$, $\ddot{Z}=0$. a) Non-stretch Rolie-Poly, b)
    Stretch Rolie-Poly ($\tau_{\rm R}=0.0316, f=0.0$), c) Stretch
    Rolie-Poly with finite extensibility, $f=10^{-4}$. Red line
    in a) shows analytical prediction for divergence in $\delta a$.}
\label{fig:rolie}
\end{figure}

If the time-dependence of the base state were disregarded, instability
to necking would correspond to any eigenvalue of this matrix having
positive real part. However its entries 
$Z,X,\tensile,T$ refer to the time-evolving base state, rendering the
eigenvalues time-dependent: indeed, these can start out all negative
before one later goes positive. Numerics and analytics (not shown)
reveal this sign change to correspond to excellent approximation to
onset of downward curvature $\ddot{Z}<0$ in the time (and so strain)
evolution of the base state $Z(t)$ and also (apart from the small
discrepancy seen in Fig.~\ref{fig:OGF}) of the tensile stress,
$\ddot{\tensile}<0$. This strongly suggests that necking will arise in
any regime of $\ddot{Z}<0,\ddot{\tensile}<0$. Some caution is needed,
however, because the eigenvectors also evolve. To examine rigorously
the onset of necking, therefore, we explicitly integrate the
linearised equations numerically. (Nonlinear effects must eventually
become important, but when depends on the size of the seeding
perturbation, which is not specified.)  Resulting contour maps of
$\delta a(t)/\delta a(0)$ are shown in
Figs.~\ref{fig:OGF},~\ref{fig:rolie}, with the criteria $\ddot{Z}=0,
\ddot{\tensile}=0,\dot{F}=0$ for comparison.

In the Oldroyd B model, Fig.~\ref{fig:OGF}a, $\delta a(t)$ grows
without bound for $\edotbar < 1/2$, albeit slowly, predicting that
necking should eventually arise in any experiment of long enough
duration, even causing filament failure unless mitigated by nonlinear
effects. For $\edotbar > 1/2$, in contrast, $\delta a(t)$ grows only
weakly before quickly saturating, signifying stability against
necking. This is consistent with the divergence in the underlying
constitutive curve: for $\edotbar>1/2$ the base state $Z(t)$ and
$\tensile(t)$ diverge in time with ever upward curvature and the
eigenvalue remains negative. (For $\edotbar<1/2$, $Z(t)$ and
$\tensile(t)$ curve down towards steady state.)

The Giesekus and fene-P models avoid this viscosity divergence: in the
evolution of the base state an early time regime of Oldroyd B-like
upward curvature $\ddot{\tensile}>0,\ddot{Z}>0$, inside the
nose-shaped region in Figs.~\ref{fig:OGF}b,c, later gives way to
downward curvature toward a high viscosity steady state.  Surprisingly
$\delta a$ does grow inside the nose, even though $\ddot{Z}>0$ and all
eigenvalues are negative here. This growth is however weak enough to
go unnoticed, assuming the perturbation that initially seeds the
instability to be small. Rapid exponential growth and observable
necking set in only after the nose, where
$\ddot{Z}<0,\ddot{\tensile}<0$ and an eigenvalue is positive.

Can these contour maps be understood analytically? Given an evolution
equation $\partial_t\tensile=\edot G(\tensile, T)-H(\tensile,T)/\tau$
for the base state's tensile stress, one can show (approximately) for
Oldroyd B and Giesekus that $\delta a(t)/\delta a(0) = f(x)$ with
$f=\left(1+b\,x\right)^{\edotbar/b}$ and
$x=\tensile/\dot{\tensile}$. Here $b=H/\tau\tensile-\edotbar$ varies
only mildly compared to $x$ in Giesekus, and is constant for Oldroyd
B. In any regime of finite viscosity (Oldroyd B for $\edotbar<1/2$ and
Giesekus for all $\edotbar$) the argument $x\to\infty$ as $t\to\infty$
due to the downward curvature to steady state $\dot{\tensile}\to
0$. Combined with $f(x)\to\infty$ as $x\to\infty$, this gives $\delta
a\to \infty$ as $t\to\infty$: necking fluctuations grow without bound,
as seen numerically. In contrast the divergent viscosity of Oldroyd B
for $\edotbar>1/2$ gives $x\to {\rm const.}$ as $t\to\infty$ and
$\delta a$ saturates.

We now use these findings in the above phenomenological (but widely
used) models as pedagogical backdrop to the more realistic Rolie-Poly
model, focusing particularly on the dramatic role of polymeric chain
stretch.

Without chain stretch, $\tau_{\rm R}=0$, the result $\delta
a(t)=\delta a(0)f(x)$ with $f=\left(1+b\,x\right)^{\edotbar/b}$,
$x=\tensile/\dot{\tensile}$ and $b=1/\tau-\edotbar$ holds
exactly. Finite extensional viscosity further gives $x\to\infty$ as
$t\to\infty$ for all $\edotbar$, and so unbounded growth of $\delta
a$. In fact for $\edotbar >1/\tau$, $f$ diverges at finite $x$ giving
a dramatic finite time divergence in $\delta a$. An analytical
prediction for this is shown in Fig.~\ref{fig:rolie}a; the numerical
contours indeed accumulate at it. Clearly, suppression of chain
stretch confers spectacular necking instability.

For finite $\tau_{\rm R}$ the base state shows divergent extensional
viscosity for $\edotbar \to 1/\tau_{\rm R}$, associated with the
development of infinite chain stretch. This is reminiscent of Oldroyd
B for $\edotbar \to 1/2$. Correspondingly, necking is suppressed
altogether for $\edotbar>1/\tau_{\rm R}$ (Fig.~\ref{fig:rolie}b). Even
for $\edotbar < 1/\tau_{\rm R}$ the finite-time divergence of
Fig.~\ref{fig:rolie}a disappears. Chain stretch thus strongly
mitigates necking.

Although instructive, infinite chain stretching is clearly unphysical
so we now introduce a parameter $f$ to cure
it~\cite{likhtmangraham03}, analogous to $\delta$ in fene-P.  With
this, the regime of upward curvature in the base state, $\ddot{Z}>0$,
$\ddot{\tensile}>0$, is again confined to a nose-shaped region
(Fig.~\ref{fig:rolie}c). As in Giesekus and fene-P, $\delta a$ grows
only slowly inside this: exponential growth sets in only beyond it,
once $\ddot{Z}<0$, $\ddot{\tensile}<0$. The nose's tip at $\edotbar
\tau_{\rm R}=O(1)$ gives a maximum in onset strain {\it vs.}
$\edotbar$ that closely resembles experimental
data~\cite{vinogradov1975c,wang}, as collated onto a master plot in
Ref.~\cite{reviews1}.

Taken together, Figs.~\ref{fig:OGF} and~\ref{fig:rolie} tell us how
heterogeneity grows in a necking instability intrinsic to the
material's rheology. To specify the process fully also requires
knowledge of what initial perturbation seeds the instability.  If this
is small (thermal noise or minor mechanical imperfection), many
contour lines will need to be crossed before necking becomes apparent:
well into the regime of exponential growth in $\delta a(t)$, and so of
$\ddot{Z}<0, \ddot{\tensile}<0$, which we thus propose as the
criterion for onset in this case.  This is likely in rheometers that
co-thin their endplates as the sample extends. In others, much larger
heterogeneous seeding will arise trivially: because the sample is
constrained to remain thicker at the plates.  Only the first few
contours might then need to be crossed and, as seen in the maps, the
\Considere criterion $\dot{f}<0$ should perform tolerably.  In either
case we expect such end-plate effects, however small, to select out of
the $q-$independent spectrum of (\ref{eqn:linear}) those modes
favouring a single neck, mid-sample.

Finally we ask if {\em any} fluid of finite extensional viscosity
should neck, via a simplified approach in which (i) the base state has
already attained a steady state on its constitutive curve; and (ii)
$Z=Z(\edot), X=X(\edot)$ are instantaneously prescribed by $\edot$.
(\ref{eqn:linear}) then reduces to $ \partial_t \delta
a=-\delta\edot$, $0=\tensile(\edotbar)\delta
a+\tensile'(\edotbar)\delta \edot$, and so $\partial_t \delta
a=(\tensile/\tensile') \delta a$. This indeed suggests any regime of
$\tensile'(\edotbar)>0$ is unstable to necking (including any
Newtonian fluid, even without surface tension).  The physics is clear.
Mass balance $\partial_t \delta a=-\delta\edot$ makes a more strongly
stretched section of filament get thinner.  This must then develop
larger stress $\tensile$ to maintain uniform force along the
thread. For positive $\tensile'$ this needs even stronger stretching,
giving positive feedback. This also suggests any {\em negatively}
sloping part of a constitutive curve to be {\em stable} against
necking. Whether any such a curve exists, and whether this distilled
result still holds given realistic $Z,X$ dynamics, is unclear.

In summary, we have given a new criterion for instability to necking
of a filament of polymeric fluid in uniaxial extension of constant
rate.  Future work will consider the nonlinear dynamics of necking,
beyond the linear regime; effects of surface tension; applied force
protocols and non-uniform extension rates; and the interplay of
``intrinsic'' necking instability with ``external'' (endplate)
effects.

The author thanks Gareth McKinley, Mike Cates, Ron Larson and Helen
Wilson for interesting discussions.


\end{document}